\documentclass[fleqn,twoside]{article}
\usepackage[headings]{espcrc2}

\readRCS
$Id: espcrc2.tex,v 1.2 2004/02/24 11:22:11 spepping Exp $
\usepackage{graphicx}
\usepackage[figuresright]{rotating}

\newcommand{\AmS}{{\protect\the\textfont2
  A\kern-.1667em\lower.5ex\hbox{M}\kern-.125emS}}

\title{$\eta^\prime$ gluonic content and $J/\psi\to VP$ decays}

\author{R. Escribano
\address[IFAE]{Grup de F\'{\i}sica Te\`orica and IFAE, Universitat Aut\`onoma de Barcelona, \\ 
                            E-08193 Bellaterra (Barcelona), Spain}%
\thanks{
UAB-FT-652.
I would like to express my gratitude to the PHIPSI08 Organizing 
Committee for the opportunity of presenting this contribution, and for the 
pleasant and interesting workshop we have enjoyed.
This work was supported in part by the Ramon y Cajal program,
the Ministerio de Educaci\'on y Ciencia under grant FPA2005-02211,
the EU Contract No.~MRTN-CT-2006-035482, ``FLAVIAnet'',
the Spanish Consolider-Ingenio 2010 Programme CPAN (CSD2007-00042), and
the Generalitat de Catalunya under grant 2005-SGR-00994.}
}
       
\runtitle{$\eta^\prime$ gluonic content and $J/\psi\to VP$ decays}
\runauthor{R. Escribano}

\begin{document}

\begin{abstract}
The $\eta$-$\eta^\prime$ pseudoscalar mixing angle and the gluonium content of the $\eta^\prime$ meson are deduced from an updated phenomenological analysis of $J/\psi$ decays into a vector and a pseudoscalar meson.
The values for the mixing angle and the gluonic content of the $\eta^\prime$ wave function are
$\phi_P=(44.5\pm 4.3)^\circ$ and $Z^2_{\eta^\prime}=0.28\pm 0.21$, respectively.
\vspace{1pc}
\end{abstract}

\maketitle

\section{INTRODUCTION}

Recently, the KLOE Collaboration reported a new measurement of the
$\eta$-$\eta^\prime$ pseudoscalar mixing angle and the gluonium content of the $\eta^\prime$ meson
\cite{Ambrosino:2006gk}.
Combining the value of $R_{\phi}$ with other constraints, they estimated the gluonium content in the
$\eta^{\prime}$ wave function as $Z_{\eta^\prime}^2=0.14\pm 0.04$
and the mixing angle in the quark-flavour basis as $\phi_P = (39.7\pm0.7)^{\circ}$.

Later on, we performed a phenomenological analysis of radiative 
$V\to P\gamma$ and $P\to V\gamma$ decays,
with $V=\rho, K^\ast, \omega, \phi$ and $P=\pi, K, \eta, \eta^\prime$,
aimed at determining the gluonic content of the $\eta$ and $\eta^\prime$ wave functions
\cite{Escribano:2007cd}.
We concluded that the current experimental data on $VP\gamma$ transitions indicated within our model
a negligible gluonic content for the $\eta$ and $\eta^\prime$.
In particular, accepting the absence of gluonium for the $\eta$, the gluonic content of the 
$\eta^\prime$ wave function amounts to $Z_{\eta^\prime}^2=0.04\pm 0.09$ or
$|\phi_{\eta^\prime G}|=(12\pm 13)^\circ$ and the $\eta$-$\eta^\prime$ mixing angle is found to be
$\phi_P=(41.4\pm 1.3)^\circ$.

Motivated by the discrepancy of the two former analyses,
we perform a new phenomenological analysis of $J/\psi\to VP$ decays to confirm or refute the possible
gluonium content of the $\eta^\prime$ meson.
This is now feasible in view of the new experimental data at our disposal which makes of the
$J/\psi\to VP$ decays the most precise and complete set of measurements
(after the $VP\gamma$ transitions) to further test this possibility.

These decays have been studied in the literature (see for instance
Refs.~\cite{Bramon:1997mf,Feldmann:1998vh,Escribano:2002iv,Li:2007ky,Thomas:2007uy}
and references therein).
In Ref.~\cite{Bramon:1997mf}, the value of the $\eta$-$\eta^\prime$ mixing angle was deduced from this relevant set of $J/\psi\to VP$ decay data including for the first time corrections due to non-ideal
$\omega$-$\phi$ mixing.
These corrections turned out to be crucial to find $\phi_P=(37.8\pm 1.7)^\circ$, 
which was appreciably less negative than previous results coming from similar analyses.
From the same set of data, a recent determination of the mixing angle assuming no gluonium finds
$\phi_P=(40\pm 2)^\circ$ while with gluonium in the $\eta^\prime$ gives $\phi_P=(45\pm 4)^\circ$ and
$\cos\phi_{\eta^\prime G}=0.84^{+0.10}_{-0.14}$ \cite{Thomas:2007uy}.

\section{NOTATION}

\label{notation}
We work in a basis consisting of the states \cite{Rosner:1982ey}
\begin{eqnarray}
\label{quarkstates}
\lefteqn{|\eta_q\rangle\equiv\frac{1}{\sqrt{2}}|u\bar u+d\bar d\rangle\ ,\qquad
|\eta_s\rangle\equiv |s\bar s\rangle\ ,}\\[1ex]
\label{gluoniumstate}
\lefteqn{|G\rangle=|\mbox{gluonium}\rangle\ .}
\end{eqnarray}
The physical states $\eta$ and $\eta^\prime$ are assumed to be the linear combinations
\begin{eqnarray}
\label{physicaleta}
\lefteqn{|\eta\rangle=X_\eta |\eta_q\rangle+Y_\eta |\eta_s\rangle+Z_\eta |G\rangle\ ,}\\[1ex]
\label{physicaletap}
\lefteqn{|\eta^\prime\rangle=
X_{\eta^\prime}|\eta_q\rangle+Y_{\eta^\prime}|\eta_s\rangle+Z_{\eta^\prime}|G\rangle\ ,}
\end{eqnarray}
with $X_{\eta (\eta^\prime)}^2+Y_{\eta (\eta^\prime)}^2+Z_{\eta (\eta^\prime)}^2=1$.
A significant gluonic admixture in a state is possible only if
$Z_{\eta (\eta^\prime)}^2=1-X_{\eta (\eta^\prime)}^2-Y_{\eta (\eta^\prime)}^2>0$.
The implicit assumptions in Eqs.~(\ref{physicaleta},\ref{physicaletap}) are the following:
i) no mixing with $\pi^0$ ---isospin symmetry, and
ii) no mixing with radial excitations or $\eta_c$ states.
Assuming the absence of gluonium for the $\eta$,
the coefficients $X_{\eta (\eta^\prime)}$, $Y_{\eta (\eta^\prime)}$ and $Z_{\eta (\eta^\prime)}$
are described in terms of two angles (see Ref.~\cite{Escribano:2007cd} for details),
\begin{eqnarray}
\label{Xetaetap}
\lefteqn{X_\eta=\cos\phi_P\ ,\qquad X_{\eta^\prime}=\sin\phi_P\cos\phi_{\eta^\prime G}\ ,}\\[1ex]
\label{Yetaetap}
\lefteqn{Y_\eta=-\sin\phi_P\ ,\hspace{1.5em} Y_{\eta^\prime}=\cos\phi_P\cos\phi_{\eta^\prime G}\ ,}\\[1ex]
\label{Zetaetap}
\lefteqn{Z_\eta=0\ ,\hspace{4.5em}  Z_{\eta^\prime}=-\sin\phi_{\eta^\prime G}\ ,}
\end{eqnarray}
where $\phi_P$ is the $\eta$-$\eta^\prime$ mixing angle and
$\phi_{\eta^\prime G}$ weights the amount of gluonium in the $\eta^\prime$ wave-function.
The standard picture (absence of gluonium) is realized with $\phi_{\eta^\prime G}=0$.

\section{EXPERIMENTAL DATA}

We use the most recent experimental data available for $J/\psi\to VP$ decays taken from
Ref.~\cite{Yao:2006px}.
The data for the $K^{\ast +}K^-+\mbox{c.c.}$, $K^{\ast 0}\bar K^0+\mbox{c.c.}$,
$\rho\eta$ and $\rho\eta^\prime$ channels remain the same since 1996 \cite{Barnett:1996hr}
and were reported by the DM2 \cite{Jousset:1988ni} and Mark III \cite{Coffman:1988ve} Collaborations.
The new measurements come from the BES Collab.,
Ref.~\cite{Ablikim:2005pr} for $\omega\pi^0$, $\omega\eta$ and $\omega\eta^\prime$ and
Ref.~\cite{Ablikim:2004hz} for $\phi\eta$, $\phi\eta^\prime$ and the upper limit of $\phi\pi^0$,
and the BABAR Collab.~for $\omega\eta$ \cite{Aubert:2006jq}.
The BES data are based on direct $e^+e^-$ measurements,
$e^+e^-\to J/\psi\to\omega\pi^0, \omega\eta, \omega\eta^\prime$ for channels involving $\omega$, 
$e^+e^-\to J/\psi\to\mbox{hadrons}$ for $\phi$, and
the $e^+e^-\to J/\psi\to\phi\gamma\gamma$ for the upper limit of $\phi\pi^0$.
The BABAR data is obtained with the initial state radiation method to lower the center-of-mass energy to the production ($J/\psi$) threshold.
Special attention is devoted to the case of $\rho\pi$.
Four new contributions have been reported since the old weighted average
$B(J/\psi\to\rho\pi)=(1.28\pm 0.10)\%$ \cite{Barnett:1996hr},
$(2.18\pm 0.19)\%$ from $e^+e^-\to\pi^+\pi^-\pi^0\gamma$ \cite{Aubert:2004kj},
$(2.184\pm 0.005\pm 0.201)\%$ from $e^+e^-\to J/\psi\to\pi^+\pi^-\pi^0$ \cite{Bai:2004jn},
$(2.091\pm 0.021\pm 0.116)\%$ from $\psi(2S)\to\pi^+\pi^-J/\psi$ \cite{Bai:2004jn}
---the weighted mean of these two measurements is $(2.10\pm 0.12)\%$, and
$(1.21\pm 0.20)\%$ from $e^+e^-\to\rho\pi$ \cite{Bai:1996rd}.
Thus, the new weighted average is $(1.69\pm 0.15)\%$ with a confidence level of 0.001
\cite{Yao:2006px}.
In Table \ref{tableB}, we show the branching ratios for the different decay channels
according to the present-day values \cite{Yao:2006px} (second column).

\begin{table*}[t]
\caption{
Experimental $J/\psi\to VP$ branching ratios (in units of $10^{-3}$) from
Ref.~\cite{Yao:2006px}.
The results of the fits with $x=0.81\pm 0.05$ and $\phi_V=(3.2\pm 0.1)^\circ$ corresponding to
Fit 1 (gluonium not allowed) and Fit 2 (gluonium allowed in $\eta^\prime$)  are also shown.}
\begin{tabular}{cccc}
\hline\\[-2ex]
Channel & Exp. & Fit 1 & Fit 2 \\[0.5ex]
\hline\\[-2ex]
$\rho\pi$ 						& $16.9\pm 1.5$		
							& $17.0\pm 1.2$ 			& $16.7\pm 1.3$ \\[0.5ex]
$K^{\ast +}K^-+\mbox{c.c.}$		& $5.0\pm 0.4$			
							& $5.2\pm 0.5$				& $5.3\pm 0.6$ \\[0.5ex]
$K^{\ast 0}\bar K^0+\mbox{c.c.}$ 	& $4.2\pm 0.4$			
							& $4.4\pm 0.5$				& $4.5\pm 0.6$ \\[0.5ex]
$\omega\eta$ 					& $1.74\pm 0.20$		
							& $1.59\pm 0.13$			& $1.53\pm 0.25$ \\[0.5ex]
$\omega\eta^\prime$ 			& $0.182\pm 0.021$		
							& $0.185\pm 0.064$			& $0.183\pm 0.353$ \\[0.5ex]
$\phi\eta$ 					& $0.74\pm 0.08$		
							& $0.66\pm 0.13$			& $0.67\pm 0.20$ \\[0.5ex]
$\phi\eta^\prime$ 				& $0.40\pm 0.07$		
							& $0.38\pm 0.10$			& $0.39\pm 0.40$ \\[0.5ex]
$\rho\eta$ 					& $0.193\pm 0.023$ 	
							& $0.206\pm 0.021$			& $0.199\pm 0.037$ \\[0.5ex]
$\rho\eta^\prime$ 				& $0.105\pm 0.018$ 	
							& $0.116\pm 0.014$			& $0.106\pm 0.037$ \\[0.5ex]
$\omega\pi^0$ 					& $0.45\pm 0.05$		
							& $0.39\pm 0.03$			& $0.43\pm 0.05$ \\[0.5ex]
$\phi\pi^0$ 					& $<0.0064$\ C.L.~90\%	
							& $0.0011\pm 0.0001$		& $0.0012\pm 0.0002$ \\[0.5ex]
\hline\\[-2ex]
\end{tabular}
\label{tableB}
\end{table*}

\section{PHENOMENOLOGICAL MODEL}

Since the $J/\psi$ meson is an almost pure $c\bar c$ state, its decays into $V$ and $P$ are 
Okubo-Zweig-Iziuka (OZI) rule-suppressed and proceed through a three-gluon annihilation diagram and an electromagnetic interaction diagram.
Aside from the OZI-suppressed diagrams common to all hadronic $J/\psi$ decays,
the doubly disconnected diagram,
where the vector and the pseudoscalar exchange an extra gluon,
is also expected to contribute to the $J/\psi$ decays.
The doubly disconnected diagram representing the diagram connected to a pure glueball state is also considered.

The amplitudes for the $J/\psi\to VP$ decays are expressed in terms of an
$SU(3)$-symmetric coupling strength $g$ (SOZI amplitude)
which comes from the three-gluon diagram,
an electromagnetic coupling strength $e$ (with phase $\theta_e$ relative to $g$)
which comes from the electromagnetic interaction diagram \cite{Kowalski:1976mc},
an $SU(3)$-symmetric coupling strength which is written by $g$ with suppression factor $r$ contributed from the doubly disconnected diagram (nonet-symmetry-breaking DOZI amplitude) \cite{Haber:1985cv},
and $r^\prime$ which is the relative gluonic production amplitude.
The $SU(3)$ violation is accounted for by a factor $(1-s)$ for every strange quark contributing to $g$,
a factor $(1-s_p)$ for a strange pseudoscalar contributing to $r$,
a factor $(1-s_v)$ for a strange vector contributing to $r$ \cite{Seiden:1988rr}, and
a factor $(1-s_e)$ for a strange quark contributing to $e$.
The last term arises due to a combined mass/electromagnetic breaking of the flavour-$SU(3)$ symmetry.
This correction was first introduced by Isgur \cite{Isgur:1976hg} who analysed corrections to
$V\to P\gamma$ radiative decays through a parameter $x\equiv\mu_d/\mu_s$ which accounts for the expected difference in the $d$-quark and $s$-quark magnetic moments due to mass breaking.
The $V\to P\gamma$ amplitudes are precisely proportional to the electromagnetic contribution to the
$J/\psi\to VP$ decay, whose dominant decay occurs via $J/\psi\to\gamma\to VP$.
These results are reproduced in the $J/\psi\to VP$ amplitudes by means of the identification
$x\equiv 1-s_e$.
The general parametrization of amplitudes for $J/\psi\to VP$ decays is written in Table \ref{tableA}.
In order to obtain the physical amplitudes for processes involving $\omega$ or $\phi$ one has to incorporate corrections due to non-ideal $\omega$-$\phi$ mixing (see Notation).
\begin{table*}
\caption{
General parametrization of amplitudes for $J/\psi\to VP$ decays.}
\begin{tabular}{cc}
\hline\\[-2ex]
Process & Amplitude \\[0.5ex]
\hline\\[-2ex]
$\rho\pi$ 						& $g+e$ \\[0.5ex]
$K^{\ast +}K^-+\mbox{c.c.}$		& $g(1-s)+e(2-x)$ \\[0.5ex]
$K^{\ast 0}\bar K^0+\mbox{c.c.}$ 	& $g(1-s)-e(1+x)$ \\[0.5ex]
$\omega_q\eta$ 				& $(g+e)X_\eta+\sqrt{2}\,r g[\sqrt{2}X_\eta+(1-s_p)Y_\eta]
                                                                                             +\sqrt{2}\,r^\prime g Z_\eta$ \\[0.5ex]
$\omega_q\eta^\prime$ 			& $(g+e)X_{\eta^\prime}
                                                                 +\sqrt{2}\,r g[\sqrt{2}X_{\eta^\prime}+(1-s_p)Y_{\eta^\prime}]
                                                                 +\sqrt{2}\,r^\prime g Z_{\eta^\prime}$ \\[0.5ex]
$\phi_s\eta$ 					& $[g(1-2s)-2e x]Y_\eta
							+r g(1-s_v)[\sqrt{2}X_\eta+(1-s_p)Y_\eta]
                                                                 +r^\prime g (1-s_v)Z_\eta$ \\[0.5ex]
$\phi_s\eta^\prime$ 				& $[g(1-2s)-2e x]Y_{\eta^\prime}
                                                                 +r g(1-s_v)[\sqrt{2}X_{\eta^\prime}+(1-s_p)Y_{\eta^\prime}]
                                                                 +r^\prime g (1-s_v)Z_{\eta^\prime}$ \\[0.5ex]
$\rho\eta$ 					& $3e X_\eta$ \\[0.5ex]
$\rho\eta^\prime$ 				& $3e X_{\eta^\prime}$ \\[0.5ex]
$\omega_q\pi^0$ 				& $3e$ \\[0.5ex]
$\phi_s\pi^0$ 					& $0$ \\[0.5ex]
\hline\\[-2ex]
\end{tabular}
\label{tableA}
\end{table*}

Given the large number of parameters to be fitted, 13 in the most general case for 11 observables
(indeed 10 because there is only an upper limit for $\phi\pi^0$),
we perform the following simplifications.
First, we set the $SU(3)$-breaking contributions $s_v$ and $s_p$ to zero since they always appear multiplying $r$ and hence the products $r s_v$ and $r s_p$ are considered as second order corrections which are assumed to be negligible.
Second, we fix the parameters $x=m_{u,d}/m_s$ and the vector mixing angle $\phi_V$ to the values obtained from a recent fit to the most precise data on $V\to P\gamma$ decays \cite{Escribano:2007cd},
that is $m_s/m_{u,d}=1.24\pm 0.07$ which implies $x=0.81\pm 0.05$ and $\phi_V=(3.2\pm 0.1)^\circ$.
The value for $x$ is within the range of values used in the literature, $0.62$ \cite{Baltrusaitis:1984rz},
$0.64$ \cite{Jousset:1988ni}, $0.70$ \cite{Morisita:1990cg}, and
the $SU(3)$-symmetry limit $x=1$ \cite{Coffman:1988ve}.
The value for $\phi_V$ is in perfect agreement (magnitude and sign)
with the value $\phi_V=(3.4\pm 0.2)^\circ$ obtained from the ratio
$\Gamma(\phi\to\pi^0\gamma)/\Gamma(\omega\to\pi^0\gamma)$ and the $\omega$-$\phi$ interference in $e^+e^-\to\pi^+\pi^-\pi^0$ data \cite{Dolinsky:1991vq}.
It is also compatible with the value $\phi_V^{\rm quad}=+3.4^\circ$
coming from the squared Gell-Mann--Okubo mass formula (see Ref.~\cite{Yao:2006px}).
Finally, we do not allow for gluonium in the $\eta$ wave function,
thus the mixing pattern of $\eta$ and $\eta^\prime$ is given by the mixing angle $\phi_P$ and the coefficient $Z_{\eta^\prime}$ (see Notation).

\section{RESULTS}

We proceed to present the results of the fits.
We will also compare them with others results reported in the literature.
To describe data without considering the contribution from the doubly disconnected diagram (terms proportional to $rg$) has been shown to be unfeasible \cite{Coffman:1988ve}.
Therefore, it is required to take into account nonet-symmetry-breaking effects.
We have also tested that it is not possible to get a reasonable fit setting the $SU(3)$-breaking correction $s$ to its symmetric value, \textit{i.e.~}$s=0$.
Although it should be easily fixed from the ratio $J/\psi\to K^{\ast +}K^-/K^{\ast 0}\bar K^0$,
the value of $x$ is weakly constrained by the fit
(see Ref.~\cite{Bramon:2000qe} for the case $\phi\to K^+K^-/K^0\bar K^0$).
For those reasons, we start fitting the data with $x=0.81$  (see above) and leave $s$ free.
The vector mixing angle $\phi_V$ is for the time being also set to zero.

If gluonium is not allowed in the $\eta^\prime$ wave function,
the result of the fit gives $\phi_P=(40.4\pm 2.4)^\circ$ ---or $\theta_P=(-14.3\pm 2.4)^\circ$---
with $\chi^2/\mbox{d.o.f.}=3.6/4$, in disagreement at the $2\sigma$ level with
$\theta_P=(-19.1\pm 1.4)^\circ$ \cite{Jousset:1988ni},
$\theta_P=(-19.2\pm 1.4)^\circ$ \cite{Coffman:1988ve}, and
$\theta_P\simeq -20^\circ$ \cite{Morisita:1990cg},
but in correspondence with $\phi_P=(39.9\pm 2.9)^\circ$ \cite{Feldmann:1998vh}
and $\phi_P=(40\pm 2)^\circ$ \cite{Thomas:2007uy}.
In some analyses, $x$ is kept fixed to one since it always appears multiplying $e$ and hence also considered as a second order contribution.
In this case, our fit gives $\phi_P=(40.2\pm 2.4)^\circ$ with $\chi^2/\mbox{d.o.f.}=3.4/4$.
However, none of the former analyses include the effects of a vector mixing angle different from zero.
It was already noticed in Ref.~\cite{Bramon:1997mf} that these effects,
which were considered there for the first time, turn out to be crucial to find a less negative value of the
$\eta$-$\eta^\prime$ mixing angle.
If we take now the fitted value $\phi_V=+3.2^\circ$ (see above), one gets
$\phi_P=(40.5\pm 2.4)^\circ$ with $\chi^2/\mbox{d.o.f.}=4.2/4$ and
$\phi_P=(40.3\pm 2.4)^\circ$ with $\chi^2/\mbox{d.o.f.}=3.8/4$ for $x=0.81$ and $x=1$, respectively.
These new fits seem to refute the strong correlation between the two mixing angles found in
Ref.~\cite{Bramon:1997mf}.
One interesting feature of the present analysis is the effect produced in the fits by the new averaged value of the $\rho\pi$ branching ratio.
For instance, if $B(\rho\pi)=(16.9\pm 1.5)\%$ \cite{Yao:2006px} is replaced by its old value
$(12.8\pm 1.0)\%$ \cite{Barnett:1996hr} one gets $\phi_P=(37.7\pm 1.5)^\circ$
with $\chi^2/\mbox{d.o.f.}=8.8/4$ for $x=0.81$ and $\phi_V=3.2^\circ$, 
\textit{i.e.}~the central value and the error of the mixing angle become smaller and the quality of the fit worse.
However, this value is now in agreement with that found in Ref.~\cite{Bramon:1997mf}.

As stated, the former fits are performed assuming the absence of gluonium in $\eta^\prime$.
Now, we redo some of the fits accepting a gluonic content in the $\eta^\prime$ wave function.
For $\phi_V=0$, the values
$\phi_P=(44.8\pm 4.3)^\circ$ and $Z^2_{\eta^\prime}=0.29\pm 0.21$ with $\chi^2/\mbox{d.o.f.}=2.3/2$ and
$\phi_P=(45.0\pm 4.3)^\circ$ and $Z^2_{\eta^\prime}=0.30\pm 0.20$ with $\chi^2/\mbox{d.o.f.}=1.9/2$
are obtained for $x=0.81$ and $x=1$, respectively.
For $\phi_V=+3.2^\circ$, one gets
$\phi_P=(44.5\pm 4.3)^\circ$ and $Z^2_{\eta^\prime}=0.28\pm 0.21$ with $\chi^2/\mbox{d.o.f.}=3.0/2$ and
$\phi_P=(44.6\pm 4.3)^\circ$ and $Z^2_{\eta^\prime}=0.30\pm 0.21$ with $\chi^2/\mbox{d.o.f.}=2.6/2$, respectively.
These fits seem to favour a substantial gluonic component in $\eta^\prime$ which is, however, compatible with zero at $2\sigma$ due to the large uncertainty.
In all cases, the mixing angle and most of the other parameters are consistent with those assuming no gluonium but with larger uncertainties due to fewer constraints.
The parameter $r^\prime$ weighting the relative gluonic production amplitude is consistent with zero and has a large uncertainty.
These results are in agreement with
the values $\phi_P=(45\pm 4)^\circ$ and $Z^2_{\eta^\prime}=0.30\pm 0.21$ 
---or $\phi_{\eta^\prime G}=(33\pm 13)^\circ$---
found in Ref.~\cite{Thomas:2007uy}.

In Table \ref{tableB}, we display for completeness the results of the fits with
$x=0.81\pm 0.05$ and $\phi_V=(3.2\pm 0.1)^\circ$
for the case of accepting (Fit 2, fourth column) or not (Fit 1, third column) gluonium in the $\eta^\prime$ wave function.

\section{CONCLUSIONS}

In summary,
we have performed an updated phenomenological analysis of an accurate and exhaustive set of 
$J/\psi\to VP$ decays with the purpose of determining the quark and gluon content of the
$\eta$ and $\eta^\prime$ mesons.
The conclusions are the following.
First, assuming the absence of gluonium,
the $\eta$-$\eta^\prime$ mixing angle is found to be $\phi_P=(40.5\pm 2.4)^\circ$,
in agreement with recent experimental measurements \cite{Ambrosino:2006gk} and
phenomenological estimates \cite{Thomas:2007uy}.
Second, if gluonium is allowed in the $\eta^\prime$ wave function,
the values obtained are $\phi_P=(44.5\pm 4.3)^\circ$ and $Z^2_{\eta^\prime}=0.28\pm 0.21$
---or $|\phi_{\eta^\prime G}|=(32\pm 13)^\circ$,
which suggest within our model a substantial gluonic component in $\eta^\prime$.
Third, the inclusion of vector mixing angle effects, not included in previous analyses,
turns out to be irrelevant.
Finally, it is worth noticing that the recent reported values of $B(J/\psi\to\rho\pi)$
by the BABAR \cite{Aubert:2004kj} and BES \cite{Bai:2004jn} Collab.~are crucial to obtain a consistent description of data.

\end{document}